# Perfectly Perform Machine Learning Task with Imperfect Optical Hardware Accelerator

Jichao Fan,[1] Yingheng Tang,[1,2] Weilu Gao[1*]

[1]Department of Electrical and Computer Engineering, University of Utah, Salt Lake City, UT 84112, USA.
[2] School of Electrical and Computer Engineering, Purdue University, West Lafayette, IN 47907, USA.
[*]Corresponding author. Email: weilu.gao@utah.edu

## Abstract

Optical architectures have been emerging as an energy-efficient and high-throughput hardware platform to accelerate computationally intensive general matrix-matrix multiplications (GEMMs) in modern machine learning (ML) algorithms. However, the inevitable imperfection and non-uniformity in large-scale optoelectronic devices prevent the scalable deployment of optical architectures, particularly those with innovative nano-devices. Here, we report an optical ML hardware to accelerate GEMM operations based on cascaded spatial light modulators and present a calibration procedure that enables accurate calculations despite the non-uniformity and imperfection in devices and system. We further characterize the hardware calculation accuracy under different configurations of electrical-optical interfaces. Finally, we deploy the developed optical hardware and calibration procedure to perform a ML task of predicting the intersubband plasmon frequency in single-wall carbon nanotubes. The obtained prediction accuracy from the optical hardware agrees well with that obtained using a general purpose electronic graphic process unit.

## MAIN TEXT

## Introduction

The widespread utilization of machine learning (ML) algorithms in a variety of applications, such as computer vision (*1,2*), the discovery of new materials and biomolecules (*3,4*), and the electronic chip design (*5*), calls for the urgent need of high throughput and energy efficient hardware accelerations of the most computation-intensive general matrix-matrix multiplication (GEMM) operations in these algorithms. However, the physical limitation of electronic hardware has started to hit the bottleneck of increasing the integration density while reducing the power consumption to process millions of GEMM operations in modern ML tasks.

General-purpose optical ML hardware is emerging as new platform to accelerate GEMM operations in a highly parallel and low-power manner, thanks to the parallelism and multiplexing of photons and nearly zero static power consumption (*6*). Early works based on bulk optical components have shown the proof-of-concept demonstrations (*7 – 12*) and the rapid advancement of modern nanofabrication processes nowadays continues drastically shrinking the footprint of these optical components for large-scale integration. For example, a mesh of thermally reconfigurable Mach-Zehnder interferometers on a two-dimensional (2D) silicon photonic integrated circuit (PIC) can perform optical matrix-vector multiplication (MVM) (*13*); massive spatial multiplexing and photoelectric multiplication in three-dimensional (3D) free-space optics can enable the execution of MVM at high speeds and low energy consumption, by exploiting additional 3rd dimension for fan-out and routing (*14*); a low-power optical vector-vector dot product computation engine is implemented in spatial light modulators (SLMs) that can host millions of active devices on a 2D plane (*15*).

Despite multiple successful system demonstrations, the current employed optoelectronic devices in these systems still suffer from poor performance, such as large power consumption and slow response speed. Thus, the device innovation is the key to fully unleashing potential of general-purpose optical ML hardware, particularly when incorporating novel nanomaterials. For example, graphene has significantly improved the electro-optic modulation speed to 30 GHz (*16*); nonvolatile chalcogenide phase change materials in PICs have enabled the in-memory computing (*17*). However, the biggest challenge associated with these nanomaterials and their devices is the scalability, where the device non-uniformity and imperfection are inevitable in large-scale arrays. The critical questions on whether and how we can deploy the optical hardware with imperfect optical components to accelerate the execution of ML tasks remain.

Here, we report a general-purpose free-space optical ML accelerator constructed based on cascaded electrically programmable SLMs and demonstrate its characterization and accurate deployment to precisely perform a ML task of predicting the intersubband plasmon frequency in doped single-wall carbon nanotubes (SWCNTs). Specifically, we develop a calibration algorithm, with which the accuracy of GEMM operations is not dependent on the device non-uniformity and system imperfection. Furthermore, we explore various parameters of optical ML accelerator hardware for fast and accurate calculations. Finally, we deploy the calibrated optical GEMM (O-GEMM) hardware for executing ML models developed for the prediction of SWCNT intersubband transition frequencies from their structural parameters. The prediction accuracies obtained from a general-purpose graphic processing unit (GPU) and the O-GEMM hardware shown an excellent agreement; both have the accuracy ~81%. This demonstration opens new opportunities of employing optical ML hardware accelerators for high throughput screening and discovery of new materials (*18*).

**Results**

*Optical GEMM (O-GEMM) Hardware*

Fig. 1A displays the operation mechanism of the optical ML hardware accelerator for MVM operations, which are implemented through cascaded SLMs and a camera. The MVM operations can be then used in GEMM calculations through block matrix multiplications (*19*, *20*). Take an example of the multiplication of a 2 × 2 weight matrix **W** and a 2 × 1 input vector $\vec{v}$, the first SLM (SLM #1) encodes the information of $\vec{v}$ and the second SLM (SLM #2) encodes the information of **W**. The elements $v_1$ and $v_2$ inside $\vec{v}$ are physically represented by the optical transmittance $T_{v,11}$ and $T_{v,12}$ of a row of electro-optic (EO) modulators in SLM #1. Both $T_{v,11}$ and $T_{v,12}$ can be electrically controlled. The same information of $v_1$ and $v_2$ is also physically represented in the second row of EO modulators ($v_{21}$ and $v_{22}$), and this duplication facilitates the calibration algorithms as discussed later. Similarly, all four elements $w_{11}, w_{12}, w_{21}, w_{22}$ in **W** are physically represented by the optical transmittance $T_{w,ij}$ ($i,j$ = 1,2) of EO modulators in SLM #2.

When the collimated incoherent light passes through the cascaded SLMs #1 and #2, the output light intensity is proportional to the multiplication of the optical transmittance of corresponding

EO modulators, which fulfills the multiplication calculations in MVM operations. At the end there is a camera to capture the generated image after the incident light is regulated by two cascaded SLMs. The summation in MVM operations is done electronically by adding the readings from camera pixels, which capture the spatially modulated light intensity corresponding to the vector and matrix elements following MVM operations. For example, the addition of camera readings on the first row, $I_{d,1}$, is proportional to $w_{11}v_1 + w_{12}v_2$ and thus the first element ($o_1$) of the output vector $\vec{o}$ of $\mathbf{W}\vec{v}$. In order to represent the bipolar elements in the input vector and weight matrix with non-negative physical quantities (e.g., $T_v$, $T_w$, and $I_d > 0$), each element $v_i$ and $w_{ij}$ ($i, j = 1,2$) are represented as the difference of two positive values, such that $\mathbf{W}\vec{v} = (\mathbf{W}^+ - \mathbf{W}^-)(\vec{v}^+ - \vec{v}^-) = \vec{o}^+ - \vec{o}^-$. As a result, there are four calls of optical ML hardware, $\mathbf{W}^+\vec{v}^+$, $\mathbf{W}^-\vec{v}^-$, $\mathbf{W}^+\vec{v}^-$, $\mathbf{W}^-\vec{v}^+$, to obtain the bipolar output vector from MVM operations. In our experiment, all elements in $\mathbf{W}$ and $\vec{v}$ are in the range of [-1, 1]. Furthermore, the GEMM operations can be done in such hardware, by being decomposed into multiple MVM operations through block matrix multiplications (*19*).

Fig. 1B shows the experimental O-GEMM setup for 2 × 2 matrices. The incoherent light beam generated by a red light-emitting diode (LED) is first coupled into a fiber and then collimated through an off-axis parabolic mirror. We utilize a reflective SLM (R-SLM) and a transmissive SLM (T-SLM) to encode vectors and matrices, respectively. The polarization states of both input and output light are configured in front of and after each SLM, so that each SLM has the largest modulation of optical transmittance. There is a spatial filter, consisting of a pair of lenses and an iris, in front of the camera to remove any coherent diffraction effect that leads to the crosstalk between camera pixels and thus wrong calculations. Particularly, such crosstalk-induced calculation inaccuracy is hard to be calibrated because these errors are input-dependent. A few groups of pixels (e.g., 60 × 60 pixels on the R-SLM), which are defined as active regions, on the R-SLM are used for encoding vectors, and the accurate alignment process identifies the corresponding regions for weight matrices encoding on the T-SLM and the calculations of output vectors on the camera. This alignment process further helps reduce the cross-talks between active regions; see *Supplementary Materials Supplementary Note 1* and *Fig. S1* for more details on the experimental determination of active regions. After capturing images on the camera, the electronic readings from corresponding active regions are added together to complete MVM operations. The detailed experimental setup information can be found in *Methods and Materials*.

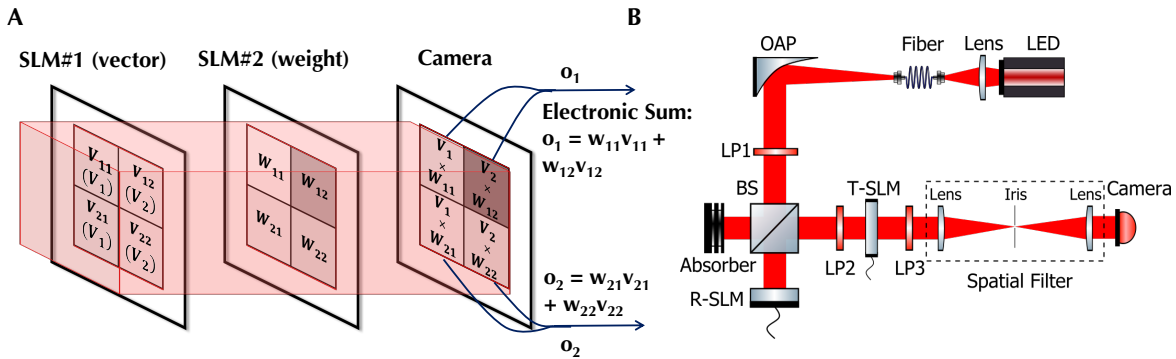

**Fig. 1. Illustration of O-GEMM principle and experimental setup.** (**A**) The mechanism of optically performing MVM operations using cascaded SLMs. (**B**) The schematic of experimental setup. LED, light-emitting diode; OAP, off-axis parabolic mirror; LP, linear polarizer; BS, beam splitter; R-SLM, reflective spatial light modulator; T-SLM, transmissive spatial light modulator.

*Calibration and Characterization of O-GEMM*

In ideal devices and systems, the response of EO modulators across the vector SLM (R-SLM) or weight SLM (T-SLM) is identical. However, this condition is extremely challenging to be achieved in practice. Instead, we create a calibration algorithm taking the inevitable non-ideality of devices and systems into account to perform accurate calculations. Specifically, we first measure modulation curves for each active region on both R-SLM and T-SLM. Note that we can only measure the response on the camera after light passing through both SLMs. Thus, in principle, we do not have access to the modulation curve of each active region. However, the manner of representing bipolar elements in vectors and matrices described before suggests that only the difference between different calls of MVM operations is crucial, which enables the accurate calculations without the precise knowledge of the modulation response of each active region. Specifically, we set gray levels on the T-SLM to have the smallest transmittance, which is not necessarily to be zero. We then sweep gray levels of each active region (e.g., $v_{11}$) on the R-SLM, while other regions (e.g., $v_{12}, v_{21}, v_{22}$) on the R-SLM are set with gray levels yielding a small transmittance. We obtain modulation curves named $f_{V,ij@wmin}$ ($i,j = 1,2$, the dashed lines in Fig. 2A) by collecting the electronic readings from corresponding camera regions; see *Supplementary Materials Fig. S2* for a detailed illustration. We repeat the same sweeping process when T-SLM is set to have the largest transmission, then obtain the modulation curves named $f_{V,ij@wmax}$ ($i,j = 1,2$, the solid lines in Fig. 2A). Similarly, the dashed and solid modulation curves in Fig. 2B for each active region on the T-SLM are obtained when the R-SLM is set to have the smallest and largest transmittance (named $f_{W,ij@vmin}$ and $f_{W,ij@vmax}$), respectively. Despite the 8-bit precision (256 available gray levels) in both SLMs, we choose the gray level range such that the modulation curves are monotonic. The resultant gray level range has 6-bit precision and is normalized to the range [0, 1].

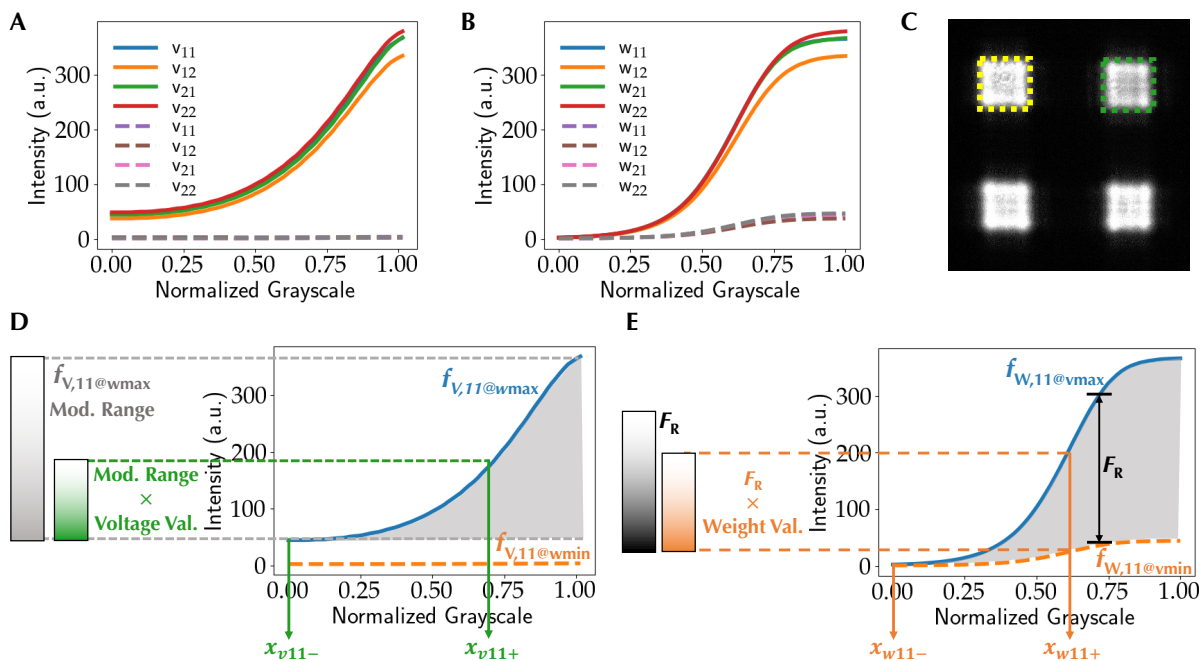

**Fig. 2. Calibration algorithm for the O-GEMM experimental setup.** The intensity modulation curves of (**A**) the R-SLM and (**B**) the T-SLM for four different matrix elements under different conditions. The grayscale range is normalized, and the intensity data is directly exported from camera readings with arbitrary unit. (**C**) One captured image on the camera showing light intensity non-uniformity. (**D**), (**E**) The illustration on how we determine gray levels for both R-SLM and T-SLM and how we utilize obtained modulation curves to perform accurate calculations.

As a result, for each pair of active regions on the R-SLM and T-SLM, there are four different modulation curves, with which we develop an algorithm to perform accurate MVM calculations with non-uniform imperfect optoelectronic devices and systems. Although mathematically $v_{11}$ ($v_{12}$) and $v_{21}$ ($v_{22}$) represent the same $v_1$ ($v_2$) in an input vector, the corresponding physical optical transmittance in active regions can be different. This physical discrepancy is nearly inevitable in practice, which can be due to the non-uniform input light beam profile and device response. Fig. 2C displays a captured image on the camera when the light passes through each pair of active regions on both SLMs. Each active region on the same SLM is under the same gray level and non-uniform brightness is clearly observed.

To explain how we perform accurate MVM calculations with obtained modulation curves, we take the first row of R-SLM and T-SLM for example. For non-negative elements (e.g., $v_{11} \geq 0$), we choose $v_{11}^+ = v_{11}$ and $v_{11}^- = 0$. For negative elements (e.g., $v_{11} < 0$), we choose $v_{11}^+ = 0$ and $v_{11}^- = |v_{11}|$. Here, we describe our calibration process in the case that $v_{11}, w_{11}, v_{12}, w_{12} \geq 0$. As shown in Figs. 2D and 2E, we can define the modulation range for the active regions on the R-SLM and T-SLM corresponding to $v_{11}$ and $w_{11}$ as $F_{R,11}$, which is expressed as $\max(f_{V,11@wmax}) + \min(f_{V,11@wmin}) - \min(f_{W,11@vmax}) - \max(f_{W,11@vmin})$. Note that $\max(f_{V,11@wmax}) = \max(f_{W,11@vmax})$, $\min(f_{V,11@wmin}) = \min(f_{W,11@vmin})$, $\min(f_{W,11@vmax}) = \max(f_{V,11@wmin})$, and $\max(f_{W,11@vmin}) = \min(f_{V,11@wmin})$. Similarly, we can define $F_{R,12}$ as $\max(f_{V,12@wmax}) + \min(f_{V,12@wmin}) - \min(f_{W,12@vmax}) - \max(f_{W,12@vmin})$. $F_{R,11}$ and $F_{R,12}$ represent the maximum tunable range obtained when light transmits through two active regions on the R-SLM and T-SLM, respectively. We define $F_{R1} = \min(F_{R,11}, F_{R,12})$ as the full tunable range of the first row, which corresponds to the case where $v_{11}, w_{11}, v_{12}, w_{12}$ are all equal to 1. Despite that $F_{R,11}$ and $F_{R,12}$ can be different due to non-uniformity, we choose the smaller one as the full range and the physical unit quantity corresponding to mathematical 1. Furthermore, such definition of $F_{R1}$ can make sure the needed tunable range for encoding $v$ and $w$ on the first row can be physically achievable in both active regions. Moreover, we can define $F_{R2} = \min(F_{R,21}, F_{R,22})$ for the second row, and $F_{R1}$ and $F_{R2}$ are completely independent thanks to separate encoding of vector information on different rows.

Without the loss of generality, we assume $F_{R1} = F_{R,12}$ to demonstrate how we obtain gray levels given input mathematical values of $v_{11} > 0$ and $w_{11} > 0$. For $v_{11}^+ = v_{11}$ and $v_{11}^- = 0$, we look for the gray level $x_{v11+}$ on the curve $f_{V,11@wmax}$ yielding $v_{11}^+ [\max(f_{V,11@wmax}) - \min(f_{V,11@wmax})] + \min(f_{V,11@wmax})$, and the gray level $x_{v11-}$ on the curve $f_{V,11@wmax}$ yielding $\min(f_{V,11@wmax})$. For $w_{11}^+ = w_{11}$ and $w_{11}^- = 0$, we first obtain a new curve $\Delta f_{W,11} = f_{W,11@vmax} - f_{W,11@vmin}$. Then, we look for the gray level $x_{w11+}$ on the curve $\Delta f_{W,11}$ yielding $w_{11}^+ F_{R1} + \min(\Delta f_{W,11})$, and the gray level $x_{w11-}$ on the curve $\Delta f_{W,11}$ yielding $\min(\Delta f_{W,11})$. Since $F_{R,11} \geq F_{R1}$ and $w_{11}^+ \leq 1$, thus $w_{11}^+ F_{R1} + \min(\Delta f_{W,11}) \leq \max(f_{V,11@wmax}) + \min(f_{V,11@wmin}) - \min(f_{W,11@vmax}) - \max(f_{W,11@vmin}) + \min(f_{W,11@vmax}) - \min(f_{W,11@vmin}) = \max(f_{W,11@vmax} - f_{W,11@vmin})$. As a result, we can always find $x_{w11+}$ to achieve desired values on $\Delta f_{W,11}$.

We collect the camera reading in a region of camera pixels corresponding to $v_{11}w_{11}$ for four times. We first set the gray level of all pixels in the $v_{11}$ active region on the R-SLM as $x_{v11+}$, and the gray level of the T-SLM as $x_{w11+}$. The average intensity in the corresponding active region on the camera, $I_{d,1}^{(1)}$, is proportional to $[v_{11}^+ (T_{v11,max} - T_{v11,min}) + T_{v11,min}] \times [w_{11}^+ F_{R1} + T_{w11,min}(T_{v11,max} - T_{v11,min})]/(T_{v11,max} - T_{v11,min})$. The multiplier corresponds to the transmittance of EO modulators on the R-SLM and the multiplicand corresponds to the transmittance of EO modulators on the T-SLM. The proportional pre-factor constant $A$ is a combination effect from the non-uniformity of light profile and detector responsivity, which is assumed to be 1 for simplicity. $T_{v11,max}$ ($T_{w11,max}$) and $T_{v11,min}$ ($T_{w11,min}$) are the maximum and minimum achievable transmittance of EO modulators in the active region corresponding to $v_{11}$ ($w_{11}$) on the R-SLM (T-SLM), respectively. Note that these values of transmittance are not accessible through experimental measurements. Similarly, we set the gray level of R-SLM as $x_{v11-}$ and T-SLM as $x_{w11+}$, and the camera reading, $I_{d,1}^{(2)}$, is proportional to $T_{v11,min} \times [w_{11}^+ F_{R1} + T_{w11,min}(T_{v11,max} - T_{v11,min})]/(T_{v11,max} - T_{v11,min})$; we set the gray level of R-SLM as $x_{v11+}$ and T-SLM as $x_{w11-}$, and the camera reading, $I_{d,1}^{(3)}$, is proportional to $[v_{11}^+ (T_{v11,max} - T_{v11,min}) + T_{v11,min}] \times T_{w11,min}$; we set the gray level of R-SLM as $x_{v11-}$ and T-SLM as $x_{w11-}$, and the camera reading, $I_{d,1}^{(4)}$, is proportional to $T_{v11,min} \times T_{w11,min}$. As a result, $I_{d,1} = I_{d,1}^{(1)} - I_{d,1}^{(2)} - I_{d,1}^{(3)} + I_{d,1}^{(4)} = v_{11}^+ (T_{v11,max} - T_{v11,min}) \times [w_{11}^+ F_{R1}/(T_{v11,max} - T_{v11,min})] = v_{11}^+ w_{11}^+ F_{R1}$. Thus, if we divide the obtained camera reading $I_{d,1}$ by the obtained $F_{R1}$ from modulation curves, we have $I_{d,1}/F_{R1} = v_{11}^+ w_{11}^+$, which yields the correct multiplication result. With this process, all non-uniformity from light profile, detector response, and SLM modulation response can be eliminated for accurate MVM operations.

To verify the calibration algorithm, we perform the MVM calculations of 1000 randomly generated matrices **W** and input vectors $\vec{v}$, where each element is uniformly randomly generated in the range [-1, 1]. The measured value of the first element of output vectors ($\widetilde{o_1}$) and the expected value obtained from standard digital computers ($o_1$) are used to define the calculation error as $(\widetilde{o_1} - o_1)/o_1$. Fig. 3A shows the scatter plot of measured and expected values, which is roughly along the line of $y = x$. The corresponding histogram plot of the calculation error distribution is shown in Fig. 3B. To have insight into those relatively large errors, we decompose the distribution for expected values in different ranges. For clarity, we only display the contribution of positive expected values. The large positive errors (cyan area in Fig. 3B) correspond to the expected values in the range [0, 0.2] (cyan dots in Fig. 3A); the golden area and dots correspond to the range [0.2, 1]; and the black area and dots cover the full value range [-2, 2]. It is clear that small values contribute to large calculation errors, since the expected value is in the denominator of error calculation, which magnifies errors. However, in neural networks, the small weights do not affect network performance much and can be pruned (*21*). As a result, the large calculation errors for small numbers are not important for accurately performing ML tasks and models. In a stark contrast, if we assume all modulation curves are identical in the ideal case and we do not perform any calibration process, the measured and expected values are completely irrelevant as shown in Figs. 3C and 3D. In addition, we intentionally increase the active region size of $v_{11}$ from 60 × 60 to 70 × 70. This uneven size in active regions can mimic the case where the fabrication yields of SLM array is not 100%, where a portion of random pixels are completely dark or unable to be modulated. As shown in Figs. 3E and 3F, the calculation accuracy is the same for the cases of even and uneven sizes after calibration.

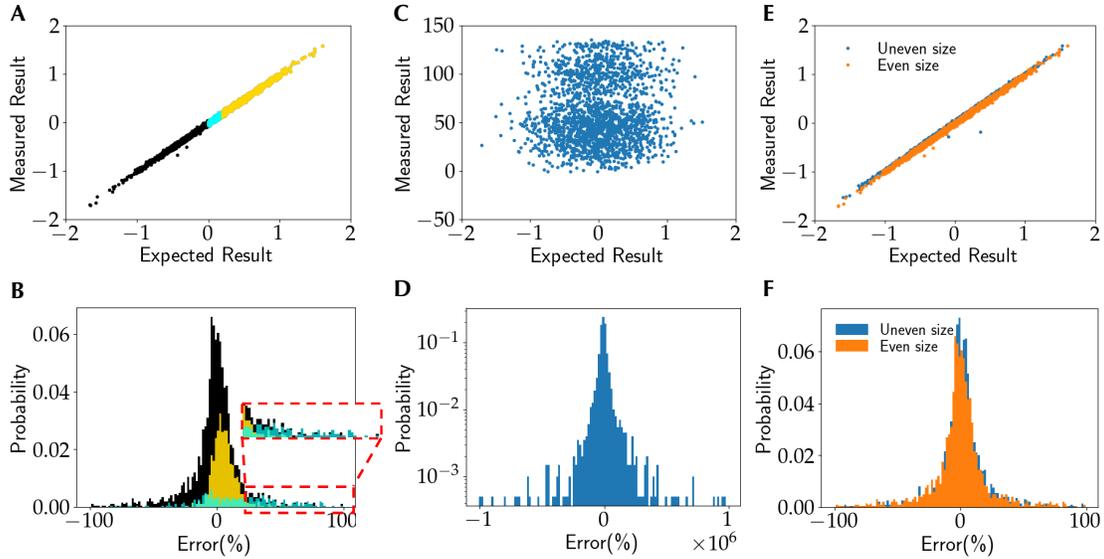

**Fig. 3. Calibration results of O-GEMM.** (**A**) Scatter plot of measured and expected multiplication results of 1000 randomly generated matrices and vectors. The measured results are obtained by using the calibration algorithm. (**B**) Corresponding error distribution. The cyan area in (**A**) and dots in (**B**) indicate the expected values in the range [0, 0.2]. Golden area and dots represent the range [0.2, 2]. Black area and dots cover the full value range [-2, 2]. (**C**) Scatter plot and (**D**) error distribution for the measurement done without the calibration algorithm. (**E**) Scatter plot and (**F**) error distribution for the measurement done under different active region size, but with the calibration algorithm.

Although the demonstrated SLMs are implemented using liquid crystal technology with slow modulation speed at 60 Hz, the high-speed (e.g., ~ 1GHz) spatial light modulation can be achieved using other technologies, such as silicon photonics (*23*) and graphene (*24*). The electrical-optical interfaces and conversions have significant influence on system speed and throughput. We explore the calculation accuracy dependence on two factors, the camera exposure time and the bit precision of analog-to-digital conversion in SLM driving voltages. The lower camera exposure time or averaging time implies faster read-out and higher throughput. Fig. 4A presents the standard deviation of 1000 MVM calculation errors as a function of exposure time. The calculation error is generally small when the exposure time is larger than 1 ms. When the exposure time goes down to 200 $\mu$s, the error significantly increases. Specifically, for 50 ms and 200 $\mu$s exposure times, Figs. 4B and 4C show the scatter plot of experimentally measured and expected values, and corresponding error distribution, respectively. Both demonstrate reduced calculation accuracy with 200 $\mu$s camera exposure time. In addition, Fig. 4D shows the error standard deviation as a function of bit precision. When the bit precision goes down to 4 bits, the accuracy substantially degrades. Furthermore, Figs. 4E and 4F show the scatter plot of experimentally measured and expected values, and corresponding error distribution for 6-bit and 4-bit precision, respectively. Both confirms the lower calculation accuracy with lower bit precision.

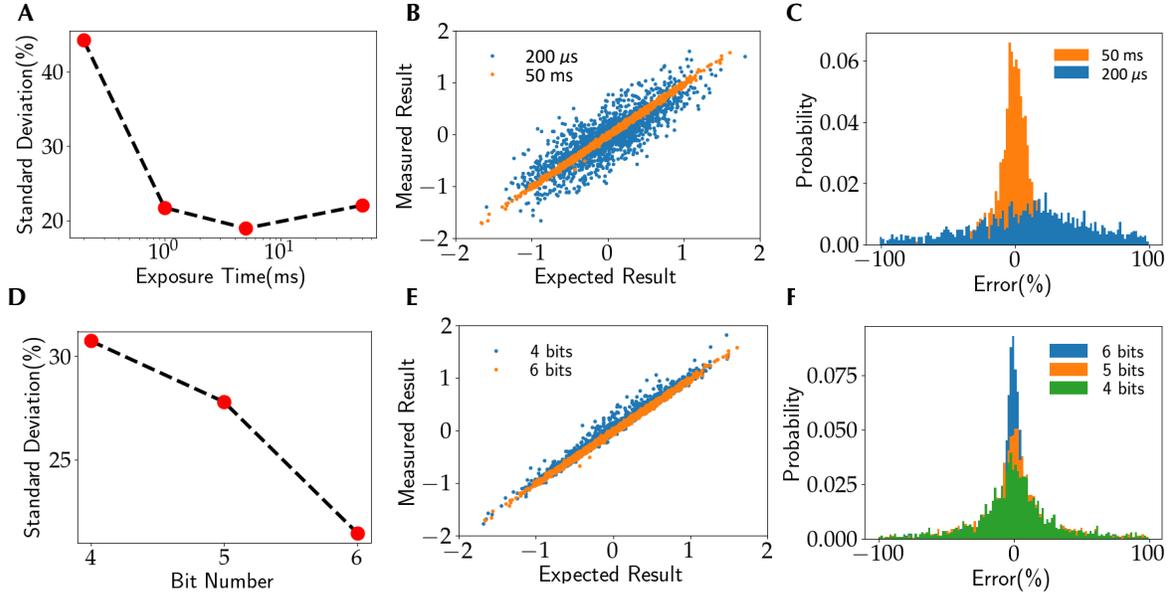

**Fig. 4. Camera exposure time and bit precision effects on accuracy.** (**A**) The standard deviation of error distribution as a function of exposure time. (**B**) Scatter plot of measured and expected multiplication results of 1000 randomly generated matrices and vectors. The measured results are obtained by using the calibration algorithm and with camera exposure time 200 μs and 50 ms. (**C**) Corresponding error distribution. (**D**) The standard deviation of error distribution as a function of bit precision, and the corresponding (**E**) scatter plot and (**F**) error distribution.

## *Accurate ML-assisted Prediction of SWCNT Intersubband Plasmons using O-GEMM*

Finally, we deploy the O-GEMM hardware setup for the inference of a ML task of predicting SWCNT intersubband plasmon frequency. SWCNTs are formed by wrapping a graphene sheet along a roll-up vector $n\vec{a_1} + m\vec{a_2}$, where $\vec{a_1}$ and $\vec{a_2}$ are primitive vectors in graphene lattice. Thus, the strong quantum confinement along the tube circumference and their one-dimensional (1D) nature lead to a variety of unique optical properties (*24*). Particularly, when the incident light is polarized perpendicular to the nanotube axis and SWCNTs are heavily doped, intersubband plasmons can be optically excited as shown in Fig. 5A (*25, 26*). Here, we consider the intersubband plasmon frequency between the first and the second subbands ($E_{12}$ in Fig. 5A), where the Fermi level $E_F$ is within the conduction band. The plasmon frequency is a function of SWCNT chirality, which are represented by integers *n* and *m*, and Fermi level. We utilize the calculated SWCNT intersubband plasmon data (*27, 28*) and divide the plasmon frequency into two classes with one greater than 1.171 eV and one smaller. Both classes roughly have 1000 data points. We construct a multi-layer perceptron (MLP) deep neural network. We have three linear layers, and the *sigmoid* nonlinear activation function is used after first two layers. A *softmax* function is used at the output; detailed network structures and input encoding are described in *Materials and Methods*.

We train the MLP model using general-purpose graphic processing unit (GPUs) and deploy the obtained network weights of linear layers on the O-GEMM hardware (Fig. 5B). The camera exposure time and the bit precision used in the O-GEMM hardware is 50 ms and 6-bit precision, respectively. Note that all the weight elements are clamped to the range [-1,1] so that the calculation can be done on the optical hardware. After the multiplication of the input vector and the weight matrix of one linear layer, the obtained output vector goes through the *sigmoid* activation function, which is executed on electronic GPUs. The *sigmoid* function regulates the output vector in the range [0,1] and the next layer GEMM calculation can also be executed on the optical hardware. Since our system has a scale of 2 × 2, the GEMM operations with the scale larger than this are done through block matrix multiplication. There are in total 2030 data, and 400 data are used for test and the rest is used for training process. Fig. 5C displays the confusion matrix on the test dataset when the GPU-trained model is inferenced on the O-GEMM hardware. The prediction accuracy is 80.7%. Fig. 5D displays the confusion matrix when the GPU-trained model is inferenced on the same GPU. The prediction accuracy is 81.2%. The performance of O-GEMM hardware after calibration is quite close to conventional electronic GPUs.

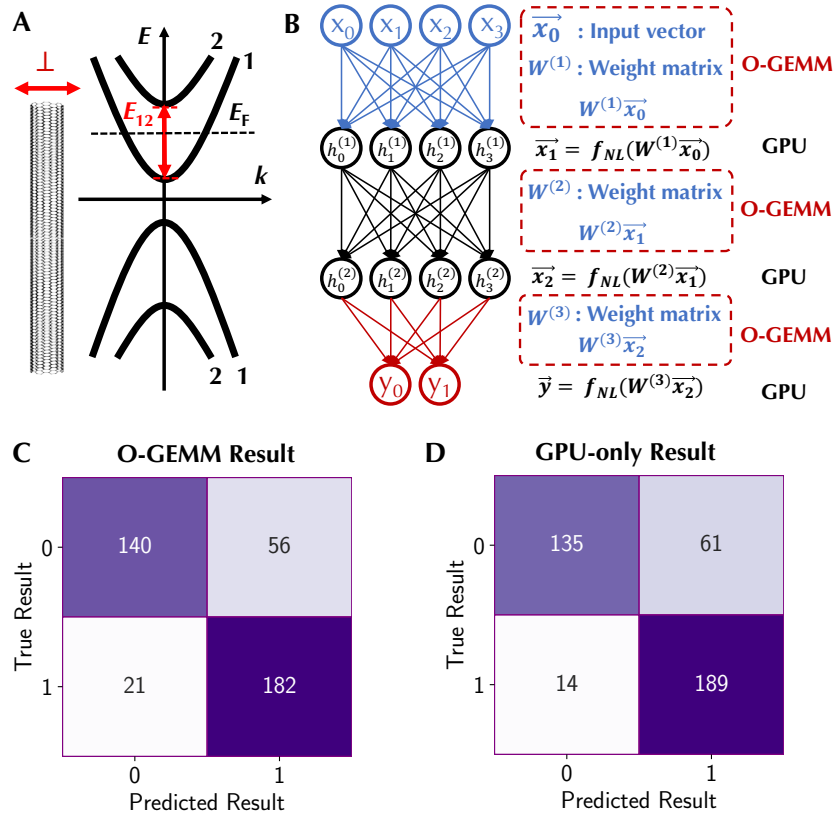

**Fig. 5. A deep neural network for predicting SWCNT intersubband plasmon frequency.** (**A**) The illustration of intersubband plasmons in heavily doped SWCNTs with the incident light is perpendicularly polarized to the tube axis. (**B**) A diagram of the inference process in the O-GEMM hardware. (**C**) and (**D**) are confusion matrices of inference results calculated using O-GEMM hardware and electronic GPUs, respectively.

## Discussion

We demonstrated a general-purpose optical hardware based on cascaded SLMs to accelerate most computation intensive GEMM operations in modern ML algorithms. Despite the inevitable imperfection of optical components and system, we developed a calibration algorithm to perform accurate calculations. The developed algorithm can be utilized in the same system architecture with any types of SLMs, particularly the SLMs based on novel nanomaterials that the imperfection in large-scale devices can be substantial. Moreover, this algorithm can be extended to other platforms, such as photonic integrated circuits, based on the similar concepts of performing MVM and GEMM operations. Thus, this calibration process lays the crucial foundation for addressing the grand scalability challenge in manufacturing optical ML systems and bridging the gap of small-scale optical ML hardware proof-of-concept demonstration to large-scale deployment in diverse disciplines.

Our current demonstration is limited to a scale of 2 × 2. The strategies of increasing the system scale include enlarging the beam size of input incoherent light to cover more pixels in SLMs by using larger optical components. Also, by replacing the halogen lamp with a laser-induced white light, we can achieve better collimation so that to reduce the spacing between active regions. Furthermore, the bright laser-induced white light and high-sensitive camera can help reduce the size of active region. With all improvement, the system scale can be as large as the SLM scale, saying ~1000 × 1000. On the other hand, the measurement of modulation curves for the calibration algorithm scales with $N^2$ for a $N \times N$ matrix, which can take significant amount of time for obtaining these curves. Faster SLMs, such as those based on the free-carrier effect in silicon-photonics-based SLMs (*22*), and high-bandwidth detector arrays can decrease the data acquisition time orders of magnitude.

**Materials and Methods**

*Experimental setup*: The LED (Thorlabs M625L4) generates red light with the center wavelength $\lambda_0 = 625$ nm and the linewidth $\Delta\lambda = 17$ nm. The generated light is coupled into a multi-mode fiber with NA = 0.39. The collimation 2-inch off-axis parabolic mirror has a focal length 152.4 mm, and the collimated beam has a diameter of 1.2 cm. A linear polarizer (LP1) has the transmission axis perpendicular to the work bench (i.e., polarization direction is perpendicular to the work bench), which is used to configure the input light polarization state for the R-SLM. A 50:50 beam splitter is placed in front of the R-SLM (Meadowlark Optics, 1920 × 1152) to route the reflective light beam to pass through the following T-SLM (HOLOEYE LC 2012). The R-SLM has the array size 1920 × 1152 with the pixel size 9.2$\mu$m × 9.2$\mu$m and the filling factor 95.7%. The T-SLM has the array size 1024 × 768 with the pixel size 36$\mu$m × 36$\mu$m and the filling factor 58.0%. A linear polarizer (LP2) is placed in front of the T-SLM to configure the output light polarization state from the R-SLM for the largest transmission modulation. At the output end of the T-SLM, there is another linear polarizer (LP3). The rotation angles of LP2 and LP3 are optimized to maximize the modulation range of T-SLM. An area of 60 × 60 pixels on the R-SLM is used to represent one element in vectors. This area roughly corresponds to an area of 20 × 20 pixels on the T-SLM, which is used to represent one element in weight matrices. We perform the search of active regions on the T-SLM to have precise alignment; see *Supporting Materials Supplementary Note 1* for more details. Before capturing spatially modulated light patterns, a 2$f$ system is used to filter out potential high-order diffraction patterns. Specifically, the 2$f$ system consists of two 2-inch lenses (focal lengths = 25 mm and 35 mm) and an iris in the middle. Images are taken by a CMOS camera (Thorlabs CS165MU). The averaged intensity in an area of 20 × 20 pixels in each bright region on the camera is used to calculate the elements in output vectors.

*Deep neural networks for SWCNT intersubband plasmons:* The SWCNT intersubband plasmon data in Ref. (*28*) consists of *n* and *m* as integers in the range of [6, 25] and [0, 14], respectively. The output plasmon frequency is in the range of [0.088, 2.406] eV. We normalize *n* and *m* by dividing by 25, which is the maximum integer for both *n* and *m*. The Fermi level is chosen from a list of [1, 1.25, 1.5, 1.75, 2] eV, which is then normalized by dividing by 2. The normalized *n*, *m*, and Fermi level form the input vector, with all elements in the range [-1, 1]. For the output labels, we choose 1.171 eV as a threshold so that the plasmon frequencies smaller than 1.171 eV are labeled as 0, while those larger than 1.171 eV are labeled as 1. The MLP neural network is composed of three dense layers, with the first layer having the input size 3 and output size 4, the second layer having the input and output sizes of 4, and the third layer having the input size of 3 and output size of 2. Between the first and second layers, as well as the second and third layers, *sigmoid* nonlinear functions are used to make sure the output from each layer is in the range that O-GEMM can execute. A *softmax* function is used at the final output end. The model is constructed, trained, and inferenced using PyTorch 1.9.1 on an Nvidia 3090 Ti GPU.

## Acknowledgements

**Author contributions:** W. G. conceived and supervised the project. J. F. constructed the experimental setup and performed all experiments under the supervision of W. G. Y. T. helped with the machine learning model. All authors discussed the manuscript. J. F. and W. G. wrote the manuscript.

**Competing interests:** The authors declare that they have no competing interests.

**Data and materials availability:** All data needed to evaluate the conclusions in the paper are presented in the paper and/or the Supplementary Materials.

## Reference

1. Y. LeCun, B. Yoshua, H. Geoffrey, Deep learning. *Nature*. **521**, 436-444 (2015).

2. I. Goodfellow, B. Yoshua, C. Aaron, "*Deep learning.*" (MIT press, Cambridge, MA, 2016).

3. K. T. Butler, D. W. Davies, H. Cartwright, O. Isayev, A. Walsh, Machine learning for molecular and materials science. *Nature*. **559**, 547-555 (2018).

4. A. W. Senior, R. Evans R, J. Jumper, J. Kirkpatrick, L. Sifre, T. Green, C. Qin, A. Žídek, A. W. Nelson, A. Bridgland, H. Penedones, Improved protein structure prediction using potentials from deep learning. *Nature*. **577**, 706-710 (2020).

5. A. Mirhoseini, A. Goldie, M. Yazgan, J. W. Jiang, E. Songhori, S. Wang, Y. J. Lee, E. Johnson, O. Pathak, A. Nazi, J. Pak, T. Andy, S. Kavya, H. William, T. Emre, V. L. Quoc, L. James, H. Richard, C. Roger, D. Jeff, A graph placement methodology for fast chip design. *Nature*. **594**, 207-212 (2021).

6. H. J. Caulfield, D. Shlomi, Why future supercomputing requires optics. *Nat. Photonics*. **4**, 261-263 (2010).

7. J. J. Hopfield, Neural networks and physical systems with emergent collective computational abilities, *Proc. Natl. Acad. Sci. U.S.A.* **79**, 2554-2558 (1982).

8. D. Psaltis, N. Farhat, Optical information processing based on an associative-memory model of neural nets with thresholding and feedback. *Opt. Lett*. **10**, 98-100 (1985).

9. T. Lu, S. Wu, X. Xu, T. Francis, Two-dimensional programmable optical neural network. *Appl. Opt.* **28**, 4908-4913 (1989).

10. G. Dunning, Y. Owechko, B. Soffer, Hybrid optoelectronic neural networks using a mutually pumped phase-conjugate mirror. *Opt. Lett*. **16**, 928-930 (1991).

11. M. Reck, A. Zeilinger, H. J. Bernstein, P. Bertani, Experimental realization of any discrete unitary operator. *Phys. Rev. Lett.* **73**, 58 (1994).

12. I. Bar-Tana, J. P. Sharpe, D. J. McKnight, K. M. Johnson, Smart-pixel spatial light modulator for incorporation in an optoelectronic neural network. *Opt. Lett*. **20**, 303-305 (1995).

13. Y. Shen, N. C. Harris, S. Skirlo, M. Prabhu, T. Baehr-Jones, M. Hochberg, X. Sun, S. Zhao, H. Larochelle, D. Englund, M. Soljačić, Deep learning with coherent nanophotonic circuits. *Nat Photonics*. **11**, 441-446 (2017).

14. R. Hamerly, L. Bernstein, A. Sludds, M. Soljačić, D. Englund, Large-scale optical neural networks based on photoelectric multiplication. *Phys. Rev. X.* **9**, 021032 (2019).

15. T. Wang T, S. Y. Ma, L. G. Wright, T. Onodera, B. C. Richard, P. L. McMahon, An optical neural network using less than 1 photon per multiplication. *Nat. Commun*. **13**, 1-8 (2022)

16. C. T. Phare, Y. H. Lee, J. Cardenas, M. Lipson, Graphene electro-optic modulator with 30 GHz bandwidth. *Nat. Photonics*. **9**, 511-514 (2015).

17. J. Feldmann, N. Youngblood, M. Karpov, H. Gehring, X. Li, M. Stappers, M. Le Gallo, X. Fu, A. Lukashchuk, A. Raja, J. Liu, C. D. Wright, A. Sebastian, T. J. Kippenberg, W. H. P. Pernice, H. Bhaskaran, Parallel convolutional processing using an integrated photonic tensor core. *Nature*. **589**, 52-58 (2021).

18. J. J. de Pablo, N. E. Jackson, M. A. Webb, L. Q. Chen, J. E. Moore, D. Morgan, R. Jacobs, T. Pollock, D. G. Schlom, E. S. Toberer, J. Analytis, I. Dabo, D. M. DeLongchamp, G. A. Fiete, G. M. Grason, G. Hautier, Y. Mo, K. Rajan, E. J. Reed, E. Rodriguez, V. Stevanovic, J. Suntivich, K. Thornton, J. C. Zhao, New frontiers for the materials genome initiative. *npj Comput. Mater.* **5**, 1-23 (2019).

19. W. Gao, C. Yu, R. Chen, Artificial intelligence accelerators based on graphene optoelectronic devices. *Adv. Photon. Res.* **2**, 210004 (2021).

20. Y. Tang *et al.,* https://doi.org/10.48550/arXiv.2203.06061 (2022)

21. S. Han, J. Pool, J. Tran, W. Dally. Learning both weights and connections for efficient neural network. *Adv. Neural Inf. Process. Syst.* **1**, 1135-1143 (2015).

22. C. Qiu, J. Chen, Y. Xia, Q. Xu, Active dielectric antenna on chip for spatial light modulation. *Sci. Rep.* **2**, 1-7 (2012).

23. C. Qiu, T. Pan, W. Gao, R. Liu, Y. Su, R. Soref, Proposed high-speed micron-scale spatial light valve based on a silicon-graphene hybrid structure. *Opt. Lett.,* **40**, 4480-4483 (2015).

24. R. B. Weisman, J. Kono, Ed., *Optical Properties of Carbon Nanotubes: A Volume Dedicated to the Memory of Professor Mildred S. Dresselhaus* (World Scientific, Singapore, 2019).

25. K. Yanagi, R. Okada, Y. Ichinose, Y. Yomogida, F. Katsutani, W. Gao, J. Kono, Intersubband plasmons in the quantum limit in gated and aligned carbon nanotubes. *Nat. Commun.* **9**, 1-7 (2018)

26. D. Satco, D. S. Kopylova, F. S. Fedorov, T. Kallio, R. Saito, A. G. Nasibulin, Intersubband plasmon observation in electrochemically gated carbon nanotube films. *ACS Appl. Electron. Mater.* **2**, 195-203 (2019).

27. D. Satco, A. R. Nugraha, M. S. Ukhtary, D. Kopylova, A. G. Nasibulin, R. Saito, Intersubband plasmon excitations in doped carbon nanotubes. *Phys. Rev. B.* **99**, 075403 (2019).

28. Data and code in Ref. 27: https://github.com/DariaSatco/cntabsorpt

# Supplementary Materials for

## Perfectly Perform Machine Learning Tasks with Imperfect Optical Hardware Accelerator

Jichao Fan, Yingheng Tang, Weilu Gao*

* Corresponding author. Email: weilu.gao@utah.edu

**Supplementary Text**

<u>Supplementary Note 1: Alignment of active regions:</u>

Four active regions on the R-SLM are first chosen around the center region of the input light beam profile. Specifically, each active region contains $60 \times 60$ pixels and the horizontal and vertical spacings between each region are 150 pixels. The relative position of beam size and active regions are illustrated in Fig. S1A. We first turn off all pixels on the R-SLM and T-SLM. We then sweep the T-SLM from the left to right, by turning on one column one time. As the result, we obtain the orange curve in Fig. S1C, which consists of the beam profile background on those closed pixels. Next, we load a two-vertical-line pattern on it with one line connecting the active regions 1 and 3 and the other connecting the active regions 2 and 4; see red dashed rectangles in Fig. S1A and the image in Fig. S1B. That means the corresponding pixels are turned on. We again sweep the T-SLM from the left to right and obtain the blue curve in Fig. S1C, which consists of contribution of open pixels and the beam profile background of closed pixels. By subtracting the orange curve from the blue curve, we obtain the green curve in Fig. S1C. From the left to the right, the first onset of rapidly increasing camera reading corresponds to the left edges of active regions 1 and 3. Then, the beginning of the first plateau corresponds to the right edges of active regions 1 and 3. Similarly, the next onset of rapidly increasing camera reading corresponds to the left edges of active regions 2 and 4, and the beginning of plateau corresponds to the right edges of active regions 2 and 4. As a result, all vertical edges of active regions are determined. By loading two-horizontal-line pattern on the R-SLM (blue dashed rectangles in Fig. S1A and the image in Fig. S1B) and performing the similar sweep procedures from the bottom to the top, we can determine the horizontal edges of active regions.

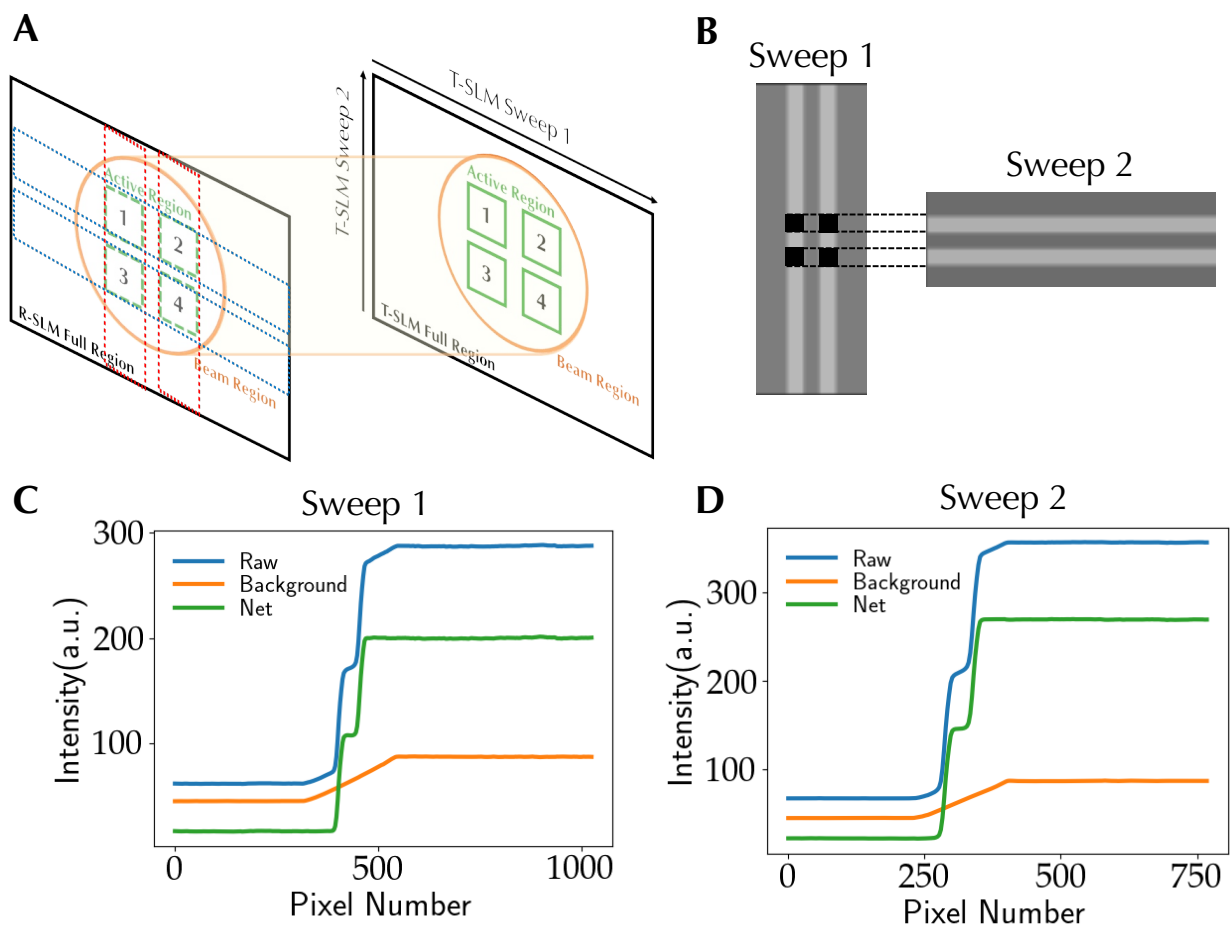

**Fig. S1. Alignment of active regions on SLMs.** (**A**)The illustration of active regions (light green squares), beam region (orange circle), and loaded line patterns (red and blue dashed rectangles) on the R-SLM, as well as corresponding regions and sweep directions on the T-SLM. (**B**) The loaded patterns on the R-SLM. (**C**) The transmitted light intensity by sweeping the T-SLM from the left to the right, when the R-SLM is dark (orange line) and is loaded with a two-vertical-line pattern (blue line). The subtraction of the orange line from the blue line is the green line. (**D**) The transmitted light intensity by sweeping the T-SLM from the bottom to the top, when the R-SLM is dark (orange line) and is loaded with a two-horizontal-line pattern (blue line). The subtraction of the orange line from the blue line is the green line.

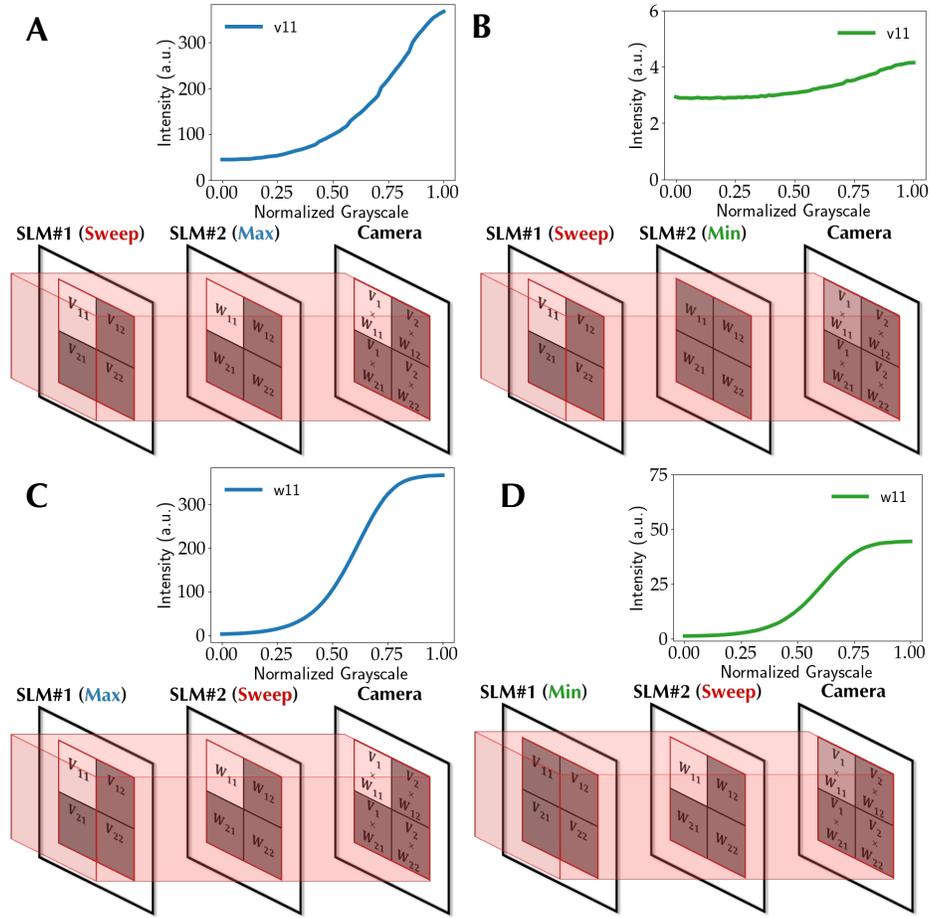

**Fig. S2. The illustration of measuring modulation curves of SLMs.** We take $v_{11}$ element for example. The light intensity modulation curves by sweeping the gray level of R-SLM (SLM #1), when the T-SLM (SLM #2) is set with (**A**) maximum and (**B**) minimum transmittance. The light intensity modulation curves by sweeping the gray level of T-SLM, when the R-SLM is set with (**C**) maximum and (**D**) minimum transmittance.

3